\documentclass[aps,pre,showpacs,twocolumn,superscriptaddress,groupedaddress,10pt]{revtex4-1}  

\usepackage{graphicx}  
\usepackage{grffile}
\usepackage{dcolumn}   
\usepackage{bm}        
\usepackage{amssymb}   
\usepackage{amsmath}
\usepackage{amsthm}
\usepackage{amsthm}
\usepackage{mathtools}
\usepackage{xcolor}
\usepackage{tikz}	
\usepackage{amsfonts}
\usepackage{calrsfs}

\usepackage[ansinew]{inputenc}

\usepackage{lipsum}

\begin{document}



\title{Unusual changeover in the transition nature of local-interaction Potts models}

\date{\today}

\author{Nir Schreiber}
\email[]{nir.schreiber@gmail.com} 
\affiliation{Department of Mathematics, Bar Ilan University, Ramat Gan, Israel 5290002}
\author{Reuven Cohen}
\affiliation{Department of Mathematics, Bar Ilan University, Ramat Gan, Israel 5290002}
\author{Simi Haber}
\affiliation{Department of Mathematics, Bar Ilan University, Ramat Gan, Israel 5290002}
\author{Gideon Amir}
\affiliation{Department of Mathematics, Bar Ilan University, Ramat Gan, Israel 5290002}
\author{Baruch Barzel}
\affiliation{Department of Mathematics, Bar Ilan University, Ramat Gan, Israel 5290002}

\begin{abstract}
A combinatorial approach is used to study the critical behavior  
of a $q$-state Potts model with a \textit{round-the-face} interaction.
Using this approach it is shown that the model exhibits a first order transition
for $q>3$. A second order transition is numerically detected for $q=2$.
Based on these findings, it is deduced that for some two-dimensional ferromagnetic
Potts models with {\it completely local} interaction, there is a changeover 
in the transition order at a critical integer $q_c\leq 3$.
This stands in contrast to the standard two-spin interaction Potts model where the maximal
integer value for which the transition is continuous is $q_c=4$.
A lower bound on the first order critical temperature is additionally derived.
\end{abstract}
 
\maketitle
 
\section{Introduction}
\label{sec1}
The ferromagnetic $q$-state Potts model \cite{Potts1952,Wu1982} is one of the most widely studied models in statistical physics. At the core of the model is an interacting spin system with each spin possessing one out of $q$ possible states. The interaction consists of the simple rule that if the  spins are monochromatic (have the same state) the energy level is lower than the case where they hold different states. 
Although the model is very easy to define, it is rich in interesting phenomena.
A particular one that has been sparsely studied is 
the changeover from a second to a first order phase transition at a critical integer value $q_c$,
or, in short, the \textit{changeover phenomenon}.

Using an equivalence of the nearest-neighbor interaction model on the square lattice
to a staggered ice-type model \cite{Temperley1971}, Baxter 
\cite{Baxter1973,0486462714} obtained an exact nonzero expression for the latent heat when $q>4$. This expression has vanished at $q=4$.
A vanishingly small jump in the free energy near $T_c$ has been found for $q\leq 4$.
Based on these results, Baxter predicted that the model underwent a continuous (discontinuous) 
transition for $q\leq q_c$ ($q>q_c$) with $q_c=4$.
His findings were believed to be lattice independent \cite{Nienhuis1979}.
Recently, Duminil-Copin and co-workers \cite{DuminilCopin2016,1611.09877}  
have rigorously confirmed Baxter's predictions 
using the random cluster 
representation \cite{Fortuin1972} of the nearest-neighbor interaction model. 

Renormalization group (RG) theory has provided a useful framework 
to detect the changeover in the transition order. 
Nienhuis {\it et\ al} considered a generalized triangular lattice Potts model with
additional lattice-gas variables and corresponding couplings,
to control order-disorder in the renormalized system.
Applying a generalized majority-rule \cite{Nienhuis1979,9780122203060}
to that model, the authors introduced a RG scheme for continuous values of $q$ which 
produced $q_c\approx 4.7$.  
Following \cite{Nienhuis1979,Nauenberg1980}, Cardy \textit{et al} \cite{Cardy1980} proposed a 
system of differential RG equations describing the scale change of the \textit{ordering, thermal, dilution} fields in the vicinity of a \textit{single} multicritical point.
The equations were derived based on the assumption that the ordering and thermal fields were relevant while the dilution field was marginal at the multicritical point.
Investigating the equations near the multicritical point, the authors have found universal, presumably exact, parameters using previously known results \cite{DenNijs1979,Nienhuis1981}.
 
It is known that some physical properties of  
all finite-range interaction models with the same dimensionality and symmetries
are universal, that is, independent of the interaction details of one model or another.
For instance, universality is manifested in the behavior of physical observables near criticality.
One can then ask whether two dimensional \textit{local} interaction ferromagnetic Potts models defined on various geometries and presenting the usual $q$-fold spontaneous symmetry breaking, in general mimic the changeover behavior of the standard pair-interaction model?  

Some variants of the pair-interaction Potts model, with long but finite range interactions \cite{Gobron2007} 
or with (exponentially) fast decaying inifinite-range interactions \cite{Chayes2009}, are known to exhibit a first order transition for $q \geq 3$. 
In another Potts model with the standard nearest-neighbor interaction, $r$ ``invisible" states were
added to the usual $q$ ``visible" states \cite{Tamura2010,vanEnter2011}. 
The invisible states affected the entropy but did not alter the internal energy. 
It has been shown 
\cite{vanEnter2011} that the transition is of first order 
for low $q$ ($q=2,3,4$), provided that $r$ is sufficiently large.

Based on Refs. \cite{Gobron2007,Chayes2009,Tamura2010,vanEnter2011}, 
the answer to the question raised above is apparently negative.
Provided undergoing a continuous transition for some $q<3$,
some of the systems \cite{Gobron2007,Chayes2009} display a changeover phenomenon at a critical
integer $q_c<3$.
It should be noted, however, that the models studied in Refs. \cite{Gobron2007,Chayes2009,Tamura2010,vanEnter2011} have a complicated  and somewhat ``unnatural" interaction content.

A phase transition entails the emergence of a monochromatic {\it giant component} (macroscopic connected component taking up a positive fraction of the sites, hereby abbreviated as GC) whose existence becomes beneficial once the energy saved by the interactions overcomes the entropy cost.  One may then ask what is the typical structure of such a monochromatic GC. In \cite{Schreiber2018} some of us used the correspondence between \textit{simple} 
 (non-fractal) and fractal-like clusters to the topology of ordered configurations in first and second order transitions, respectively, 
to show that for the round-the-face interaction Potts model on the square lattice, a changeover in 
the transition nature occurs at $q=4$.
However, Monte-Carlo (MC) simulations \cite{Schreiber2018} 
of the model with $q=4$ resulting in a
pronounced double-peaked shape of the pseudo-critical energy \textit{probability distribution function} (PDF) \cite{Janke1993}, hints 
that a first-order transition is possible then.

In the present paper we develop a more rigorous approach, based on first principles, to 
study the phase transition of the round-the-face interaction Potts model.
Using this approach, it is demonstrated on a simple ``natural" model that satisfies
the usual Potts $q$-fold broken symmetry, that within the class of $q$-state models with
completely local interactions, a changeover phenomenon occurs at
an integer value need not equal four.
More precisely, we show that the model, defined on the honeycomb lattice, undergoes a first order
transition for $q>3$. 
Under a few further conceivable assumptions related 
to the asymptotic number of hexagonal lattice animals (polyohexes)
a first order transition is detected also for $q=3$.
Extensive MC simulations \cite{Wang2001} 
are performed for different values of $q$.
For $q=2$, a second order transition is numerically observed.
It follows that the honeycomb lattice model change over its transition 
nature at a critical value $q_c\leq 3$.

The rest of the paper is organized as follows.
In section\ \ref{sec2} we discuss the role 
of large scale lattice animals in 
determining the transition nature of the round-the-face interaction model. In section\ \ref{sec3} we present and analyze in detail
the results of Monte Carlo simulations. Our conclusions are drawn in section\ \ref{sec4}.

\section{Lattice Animals and the changeover phenomenon}
\label{sec2}
Consider a ferromagnetic Potts model where the interaction involves 
$l$ spins residing at the vertices of an elementary cell (face) of a lattice with $N$ sites (e.g., $l = 4$ for the square lattice).
The model may be described by the Hamiltonian 
\begin{equation}
\label{eq:H}
-\beta{\cal H} = K\sum_{\text{faces}}\delta_{\sigma_{\text{face}}}\; ,
\end{equation}
where $\beta =1/k_B T$, $K=\beta J$ and $J>0$ is the ferromagnetic coupling constant
(we will take from now on $k_B=J=1$ for convenience). The $\delta_{\sigma_{\text{face}}}$ symbol 
assigns $1$ if {\it all} the spins $\sigma_{\text{face}} :=\{\sigma_1,\sigma_2,...,\sigma_l\}$ on a given face simultaneously take 
one of the $q$ possible Potts states, and $0$ otherwise.
The partition function of the Hamiltonian\ (\ref{eq:H}) 
\begin{equation}
\label{eq:prt_fnc}
Z = \sum_{\{\sigma\}}\prod_{\text {faces}}(1+v\delta_{\sigma_{\text{face}}})\;, 
\end{equation}
where $v = e^K-1$, can take the form \cite{Wu1982,Baxter1973,0486462714}
\begin{equation}
\label{eq:Z}
Z = q^N\sum_G q^{c(G)-\nu(G)}v^{f(G)}(1-1/q)^{\alpha(G)p(G)}\; ,
\end{equation}
where $G$ is a graph made 
of $c(G)$ clusters with a total number of $f(G)$ faces placed on its edges and $\nu(G)$ nodes. 
The total number of perimeter nodes is $p(G)$ and $\alpha(G)\le 1$ is 
a positive proportionality factor.
 
Any cluster is either a lattice animal \cite{9781498711395}
or a lattice {\it beast}, that is, a collection of animals sharing joint nodes 
such that these animals are left unconnected. 
Let $b_{km}$ be the number of beasts with $k$ faces and $m$ sites. 
It is known\ \cite{Klarner1967,Madras1999} that the total number of animals 
of size $k$ on a two-dimensional periodic lattice takes the asymptotic form $\sum_{m} b^\prime_{km}\approx c\lambda^kk^{-1}$ (the prime refers to
animal-like clusters), where $c$ is some constant.  
The total number of $k$-face beasts 
therefore asymptotically assumes $\sum_m b_{km}\sim \hat{\lambda}^k$
where $\hat\lambda\geq \lambda$. 

A beast with holes contains finite clusters with boundaries colored differently than the surrounding bulk.
In the following we refer to beasts with \textit{no} holes.
Consider a beast
with $m/k\approx \rho$ where $\rho$ is the minimal asymptotic number of sites per face (e.g., for a perfect square on the square lattice with $m = (\sqrt k+1)^2$, $\rho=1$).
Necessarily, this beast is simple.
Identifying a simple beast with a boundary 
of size $B = o(N)$, its number of sites per face is no larger than  
$(m+\Delta m)/k$ with $m/k\approx \rho$ and $\Delta m = o(N)$, hence it approaches $\rho$ in the large $N$ limit. Simple combinatorics shows that the number of simple $k$-face beasts is bounded by $KN^{a B}$ for some constants $a,K$, i.e., sub-exponential in (large) $N$.
It follows that for a given $\delta>0$ a family 
of (quasi)fractal beasts (QFs) with $m/k \approx \rho+\delta$, exponentially-growing at a rate $\mu\equiv\mu(\delta)$, is established.  
Making such a family of beasts monochromatic changes the entropy in the amount of
\begin{eqnarray}
\label{eq:real_Delta_s}
\Delta S & = & \ln\left(\mu^k/q^m\right) + \mbox{perimeter term} + \mbox{higher order terms}\nonumber \\
&=&k\ln\left(\mu/q^{\rho+\delta}\right) + \theta k\ln(1-1/q)+ o(N)\; ,
\end{eqnarray}
where $\theta k$ is a fraction	 of $O(N)$ perimeter sites.
Consequently, the change in the free energy $\Delta F = -k -T\Delta S$ (for large $k$) satisfies 
\begin{equation}
\label{eq:Delta_F}
\Delta F \geq -k(1-T\rho\ln q)-Tk\ln\left(\mu/q^{\delta}\right)\; .
\end{equation} 
Now, at $T^\ast = 1/\rho\ln q$, the temperature at which the energy gain balances the entropy loss due to the formation of 
a simple GC (effectively associated with $\delta=0,\mu=1$ and $\theta = 0$ in\ (\ref{eq:real_Delta_s}),(\ref{eq:Delta_F})), 
$\Delta F \geq   0$ if $\mu/q^\delta\leq 1.$
This means that for $q\geq q_s$, where  
\begin{equation}
\label{eq:sup}
q_s =\sup_\delta \mu^{1/\delta}\; ,
\end{equation}
it is disadvantageous 
for the system to occupy QFs at $T^\ast$.
Instead, the system possesses a simple GC at that temperature
and a first order transition is exhibited.
Conversely, if the system undergoes a second order transition then for
some $q < q_s$ and some $0<\delta\leq 2\rho$ there exist an exponential family $\mu$ such that 
$\mu/q^\delta >1$.

 
Note that in \eqref{eq:Delta_F}, a contribution  
generated by a macroscopic number of ``small" clusters is ignored.
To roughly estimate this contribution we consider the free energy of the disordered state 
$F=\langle E\rangle -TS$, where the
entropy is bounded by $N\ln q$, and the mean energy is approximated by 
$\langle E\rangle \approx -\vartheta N p_s$ where $\vartheta N$ is the total number of 
faces of the small clusters and 
\begin{equation}
\label{eq:av_E}
 p_s = \frac{qe^K}{qe^K + q^\ell-q}\; ,
\end{equation}
is the probability for a \textit{single} face to be monochromatic 
with $\ell = 2\rho + 2$ being the number of spins in an interaction (e.g., $\ell=6$ for the honeycomb lattice). In evaluating $\langle E\rangle$ one can include additional terms accounting also for monochromatic pairs of faces etc. However, these can be seen to give smaller contributions.
Adding $\langle E\rangle$ with $p_s^\ast$, the probability in \eqref{eq:av_E}, computed at $T^\ast$, or, at $e^K=q^\rho$, 
to the right-hand-side of \eqref{eq:Delta_F} results  (taking $\vartheta N = k$) in the approximated first order critical temperature 
\begin{equation}
\label{eq:Tc_1st_app}
T_c\approx \frac{1-p_s^\ast}{\rho\ln q}\sim T^\ast-\frac{1}{q^{\rho+1}}\; .
\end{equation}
In App. \ref{app1} the lower bound $\hat T=1/\ln(q^\rho+1)$ on $T_c$ is obtained.
Indeed, $|T^\ast-\hat T|\gtrsim q^{-\rho -1}$ already for small $q$.

We focus now on the honeycomb lattice since for this lattice only ``pure'' animals 
survive (there are no beasts made of animals with joint nodes). Consider an animal with no holes. 
Let $n_1,n_2$ and $n_3$ be the number of sites belonging to 
one, two and three faces, respectively. The total number of boundary sites is $b$. 
A simple counting gives
\begin{eqnarray}
\label{eq:n_123}
n_1 + 2n_2 + 3n_3 &=& 6k\; ,\nonumber\\
n_1 + n_2 + n_3 &=& m\; ,\nonumber\\
n_1 + n_2 &=& b\; .
\end{eqnarray}
One can easily verify that the minimal asymptotic number
of sites per face is $\rho=2$. Expressing $b/k$ and $\delta + o(1) = m/k-2$ in terms of $n_1,n_2,n_3$ and noticing 
that $n_1=n_2+6$ \footnote{Travelling along the perimeter, every $n_1$ vertex contributes a curvature of $\pi/3$ and every $n_2$ vertex contributes a negative curvature $-\pi/3$. Since the total curvature is $2\pi$, we have that $n_1=n_2+6$.}
 result in 
\begin{equation}
\label{eq:e/k}
b/k\leq 2\delta + o(1)\; .
\end{equation}
It is known \cite{Nienhuis1982,DuminilCopin2012} that the connective constant of self-avoiding walks (SAWs) on the honeycomb lattice is $\mu_c=\sqrt{2+\sqrt 2}$.
The exponential number of animals of size $k$ in a family with a growth constant $\mu$ is bounded by the number of SAWs of length $\tilde{b}=\sup_{n_1} b$,
i.e., $\mu^k\leq \mu_c^{\tilde{b}}$. With the help of\ (\ref{eq:e/k}) we thus obtain
\begin{equation}
\label{eq:delta_rigor}
\mu\leq \mu_c^{2\delta+o(1)}\; .
\end{equation}
Combining\ (\ref{eq:sup}) and\ (\ref{eq:delta_rigor}) together
it follows that $q_s \leq \mu_c^2\approx 3.4$. 
Consequently, the system undergoes a first order transition for $q >3$.
  
Indeed, our simulations indicate that a first order transition already occurs
at $q=3$.
Under a few plausible assumptions, this result can be analytically derived.
First, define a {\it snake} to be an animal with boundary sites only ($n_3=0$),
such that any of its faces (excluding the head and the tail) has two neighboring faces.
Next, consider {\it constrained} SAWs tracing animals in a way that allows these walks to share three edges at most with any face along their travel. 
Consider all the faces with edges mutual to a given constrained walk and placed, say, to its left.
It may occur that these faces construct a snake.
Conversely, the single side boundary of any snake can be traced by a trajectory 
overlapping with no more than three edges per face.
Thus, the number of snakes \footnote{Other animals with $n_3=0$ containing junctions, each branches into three snakes,
can be treated as if they were solely snakes, provided the number of junctions is $o(N)$.}
 is no larger than the number of 
constrained walks.
Noticing that snakes are animals associated with (maximal) $\delta = 2 $, we may write
\begin{equation}
\label{eq:mu_delta}
\mu\leq \mu_c^{\varepsilon\delta/2},\ \delta\approx 2 \; ,
\end{equation}
with $\varepsilon\leq 3$. 
Assuming that\ (\ref{eq:sup}) is governed by quasi-snakes satisfying\ (\ref{eq:mu_delta})
and substituting $\varepsilon=3$ in\ (\ref{eq:mu_delta})
imply $q_s<3$ hence a first order transition for $q\geq 3$. 

To conclude this section we point out that \eqref{eq:e/k}
holds also for the triangular and square lattices.
This is shown (see App. \ref{app3}) using counting arguments similar to\ (\ref{eq:n_123}).
It follows that since \ (\ref{eq:sup}) is 
governed by a family $\mu$ most likely dominated by animals large enough such that\ (\ref{eq:delta_rigor}) is valid, the changeover phenomenon is a universal property of the 
round-the-face model.

\section{Monte Carlo simulations}
\label{sec3}
In order to numerically support our predictions for the honeycomb lattice 
we perform Monte-Carlo simulations using the Wang-Landau \cite{Wang2001} entropic sampling method.
According to this method one hops between successive spin
configurations with energies $E_i$ and $E_j$, respectively,
with a transition probability $T_{i\to j}=\min\{1,\Omega(E_i)/\Omega(E_j)\}$
where $\Omega(E_i)$ is an approximation to the density of states with energy $E_i$. 
At each visit to a state with energy $E_i$ the corresponding quantity
$\ln\Omega(E_i)$ is modified.
We use lattices with linear sizes $5\leq L\leq 49$ 
and periodic boundary conditions 
are imposed. We focus on $q=2,3,4$. The specific heat maximum $C_L^{\max}$ is measured for each $L$
and finite size scaling (FSS) analysis is performed to these measurements.
The location (temperature, $T_L$) of $C_L^{\max}$ serves as a definition to the pseudo-critical transition temperature.

A more sensitive measure, designed for first order transitions is $T_L^W$,
the size-dependent temperature where $W_o$, the weight of the energy PDF in the ordered phase is $q$ times $W_d$, the weight in the disordered phase \cite{Janke1993}.
To compute this quantity, an initial temperature is determined such that roughly $W_o/W_d\approx q$, where $W_o,W_d$ are taken to be $\varphi(\epsilon_{\min}),1-\varphi(\epsilon_{\min})$,
respectively, and $\varphi(\epsilon_{\min})$ is the \textit{cumulative distribution function}
computed at $\epsilon_{\min}$, the energy where there is a minimum between the peaks \cite{Janke1993}.
The PDF is then iteratively calculated for increasing (decreasing) temperatures until convergence is achieved, that is, until 
\begin{equation}
\label{eq:weights}
|W_o/W_d-q|< \tilde{\epsilon}\; ,
\end{equation}
is satisfied for some $\tilde{\epsilon}$.


For both $q=3,4$, using a sample of size $L=45$ and taking $\tilde\epsilon = 5\times 10^{-3}$, a temperature increment $\Delta T = 10^{-8}$ is needed to satisfy \eqref{eq:weights}, while for $q=4$, $L=49$ and $\Delta T=10^{-8}$ convergence is attained for $\tilde{\epsilon} = 10^{-3}$. 

In Table \ref{table1} we summarize the quantities $T_L$ and $T_L^W$ together
with the bounds $\hat T = 1/\ln( q^2+1)$.
Both measured temperatures reasonably agree with $\hat T$.
Further simulations using larger samples, however, are needed
to reliably estimate $T_c$.

\begin{table}[htb!!]
\centering
\begin{tabular}{cccccc}
\hline\hline
\multicolumn{1}{c}{ } & $L$ & $T_L$ & $T_L^W$  & $\hat T$\\

\hline
$q=3$ & $45$ &  $0.4420$ & $0.4415(5)$ & $0.4342(9)$  \\ 
\hline
$q=4$ &
\begin{tabular}{c}$45$ \\ $49$\end{tabular} & \begin{tabular}{c} $0.3530$ \\ 
$0.3534$\end{tabular} & 
\begin{tabular}{c}$0.3526(9)$ \\ $0.3531(7)$\end{tabular}  & $0.3529(5)$ \\ 
 
\hline\hline
 
\end{tabular}
\caption{Finite size lattice pseudo-critical temperatures $T_L$ and $T_L^W$
together with the
lower bounds $\hat T$ on the first order critical points.}\label{table1}
\end{table}

We next fit the specific heat maxima $C_L^{\max}$ with a power-law
$C_L^{\max}\sim L^\gamma$ for $q=3,4$.
In order to prune out finite size effects and systematically detect
the FSS of large samples, we consider the maxima of the specific heat \textit{per site}
$C_L^{\max}/N,\ N=L^2$ and fit this
observable with a power-law $C_L^{\max}/N\sim L^{-\nu}$ where $\nu=d-\gamma$.

\begin{figure}[h]
\begin{center}
\hspace*{-1.4cm}
\includegraphics[width = 11cm,height = 6.5 cm]
{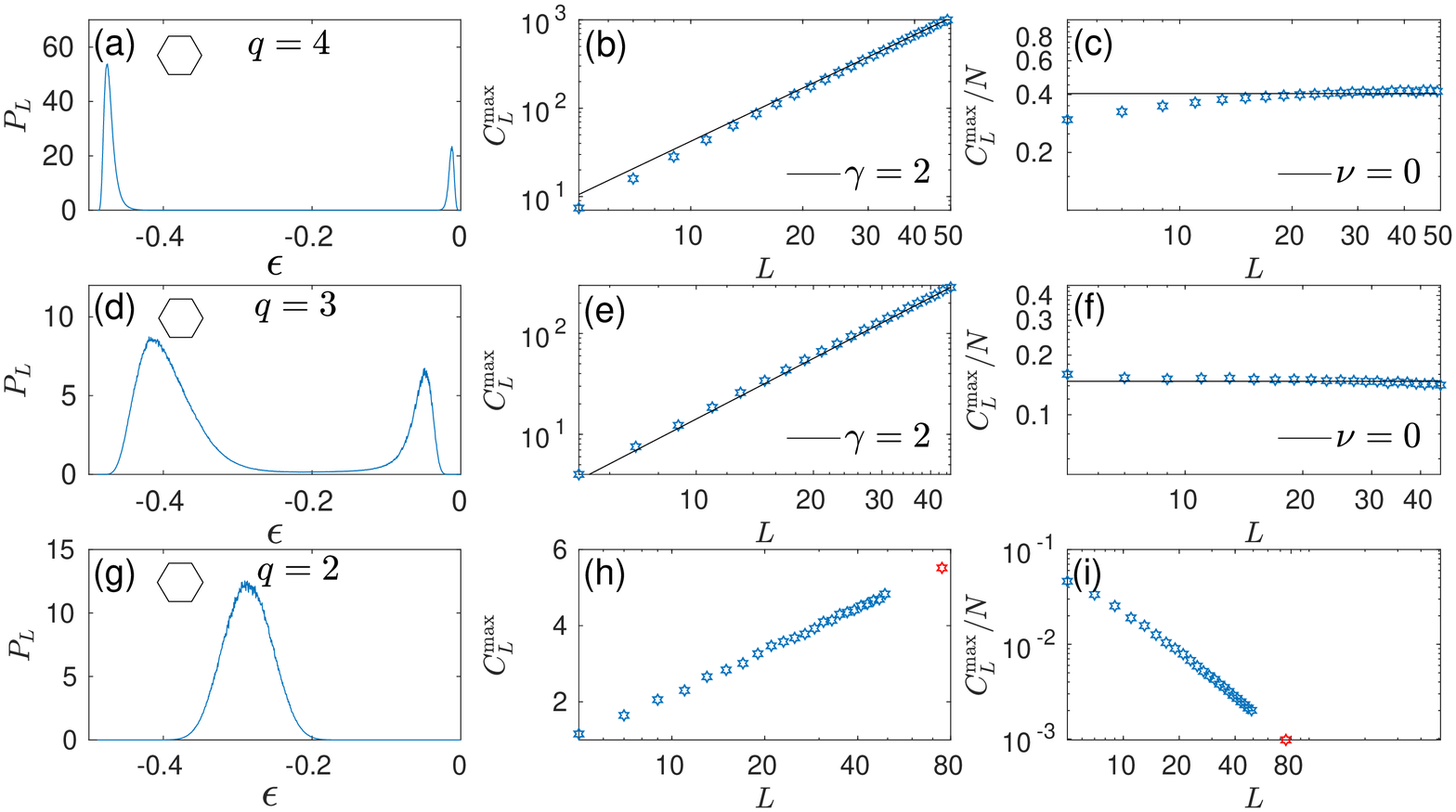} 
\end{center}
\caption{\label{fig:1}
Energy pseudo-critical PDF 
and FSS of related observables for the honeycomb lattice.
The PDF (computed for $L=45$), FSS of the specific heat maxima
and FSS of the specific heat per site maxima are
presented from left to right, respectively.
The quantities described above are associated (from top to bottom)
with  $q=4,3,2$.
The red symbols in (h),(i) represent the large sample ($L=75$) Metropolis based observables.
} 
\end{figure}

The results of the numerical analysis are graphically captured in Fig.\ \ref{fig:1}.
Fig.\ \ref{fig:1}(a) shows the PDF at $T_L^W$ for q = 4 which displays a strong first-order performance for a rather small sample ($L=45$).
In Fig.\ \ref{fig:1}(b) $C_L^{\max}\sim L^2$ apparently obey the expected first order
$C_L^{\max}\approx L^d$ scaling law,
suggesting that we use sample sizes $L$ comparable with 
the correlation length. The first order nature is supported by a slowly 
varying $C_L^{\max}/N$ (Fig.\ \ref{fig:1}(c)), indicating 
an expected first order asymptotic non-vanishing bahavior.  
Continuing  with $q=3$ and Fig.\ \ref{fig:1}(d), 
the typical pronounced double picked scenario for
the PDF at $T_L^W$ 
is clearly observed, although this quantity
apparently suffers from 
finite size effects, depicted mainly in the width of the peaks.
Fitting $C_L^{\max}$ with $L^2$ as shown in\ Fig.\ \ref{fig:1}(e), 
provides an indication for a first order behavior in the case of $q=3$. 
Similarly to the $q=4$ model, the nice fit with a zero-slope
straight line observed in Fig.\ \ref{fig:1}(f), suggests for an expected first order asymptotic constant term. The picture is substantially different for $q=2$. 
Unlike the dual peak shape characterizing the PDF of the former two models,
the current model shows (Fig.\ \ref{fig:1}(g)) 
a pronounced single peaked PDF (computed at $T_L$), indicating a continuous transition.
Motivated by this qualitative difference it seems natural to expect for
an Ising-like behavior when $q=2$, i.e., a logarithmic divergence of 
the specific heat, as captured by Fig.\ \ref{fig:1}(h).
Finally, as seen in Fig.\ \ref{fig:1}(i), 
the pseudo-singularity of the specific heat per site decays with $L$,
as expected from second order systems.
Furthermore, the rapid decay
hints that the $L^{-d}$ Ising scenario is likely to 
take place for larger samples.

In order to better apprehend the Ising-like nature of the $q=2$ model we employ the 
Metropolis \cite{Metropolis1953} method to calculate $C_L^{\max}$ and $C_L^{\max}/N$,
using the energy histogram of a sample with $L=75$. 
The associated temperature $T_L=0.6549$ is extrapolated by fitting the WL data
with the Ising form $|T_L-T_c|\propto L^{-1}$ where $T_c = 0.6596$ 
$(L_{\min}^{\rm{best}}=11,\ \chi^2/\mbox{d.o.f.}=0.04/12,\ p\ \mbox{value}=4\times 10^{-6}$ \cite{Schreiber2018,MartinMayor2007}). 
Indeed, as evident from Figs.\ \ref{fig:1}(h),\ref{fig:1}(i),
the FSS based on the WL simulations is preserved.
Further analyses of Metropolis simulations on large scale and 
WL-based energy-dependent observables are provided in App. \ref{app2}.

 
 
\section{Conclusions}
\label{sec4}
The changeover phenomenon in local-interaction ferromagnetic Potts models is studied. 
The close link between  
the round-the-face interaction Potts model
and two-dimensional lattice animals, applied to the honeycomb lattice,  
together with Monte-Carlo simulations, are used to derive $q_c\leq 3$.
Specifically, numerical results and analytical considerations indicate a first order transition 
for $q=3$. If this was to be the case, there would be a changeover phenomenon at $q_c=2$.

Our results for $q_c$ 
stand in contrast to the well known result $q_c=4$ of the usual model
\cite{Baxter1973,0486462714,1611.09877,DuminilCopin2016}.
Thus, first, by demonstrating on a natural local-interaction model a changeover phenomenon 
at a critical value $q_c<4$, 
we provide a deeper insight on this problem
strengthening the realization that $q_c=4$ being universal is a false statement \cite{Gobron2007,Chayes2009,Tamura2010,vanEnter2011}.
Second, our analytical and numerical analyses propose that 
the RG framework of Ref. \cite{Cardy1980}
assuming a single multicritical point, does not capture a changeover phenomenon in multiple models.
In other words, our study suggests that there may be multiple
multicritical points were critical and tricritical branches
terminate, keeping the thermal and ordering field relevant 
and the dilution field marginal at these points.   
Equivalently, our work may push the boundaries of the  
conventional {\it ordering field, temperature, dilution} picture \cite{Nienhuis1979,Cardy1980,20Blte17}, 
such that a new scaling field allowing for the variation of $q_c$ is introduced.
\acknowledgements

We thank the anonymous referee for useful comments and suggestions.
G.A. was supported by the Israel Science Foundation grant No. 575/16 and the German Israeli Foundation grant No. I-1363-304.6/2016.

\appendix
\setcounter{figure}{0} \renewcommand{\thefigure}{A.\arabic{figure}}

\section{A lower bound on the first order critical temperature}
\label{app1}
In the following we show that
\begin{equation}
\label{eq:Tc_low}
\hat T = \frac{1}{\ln (q^\rho+1)}\; ,
\end{equation}
is a lower bound on $T_c$,
the first order critical point.

Consider the partition function $Z_g$ 
associated with the giant component at the vicinity of $T_c$. 
Introducing the variables $f(G)=k,\ \nu(G)=m$ and taking $c(G)=1$ in \eqref{eq:Z}, $Z_g$ takes the form  
\begin{equation}
\label{eq:Z_g}
Z_g\propto q^N\sum_{m}b_{km}q^{-m}v^k\times \left(\rm perimeter\ terms\right)\; .
\end{equation}
Assuming \textit{no holes}, the number of simple beasts $\sum_m b_{km}$ is sub-exponential.
The free energy per site is then governed by the single beast exponential terms and, taking $m = \rho k + o(N)$, reads

\begin{eqnarray}
\label{eq:f_app}
        \phi&=&-T\ln\lim_{N\to \infty}Z^{1/N}\\ 
         &=&\left\{ \begin{array}{ll}
             -T\ln q + T \epsilon_o\ln (v q^{-\rho})  &, v>q^\rho \nonumber \\
             -\hat T\ln q  & , v = q^\rho\; ,
        \end{array} \right.
\end{eqnarray}
where $\epsilon_o = -\lim_{N\to\infty}k_o(N)/N$ is the energy per site which in general depends on $v$, $k_o(N)$ maximizes $Z_g$
and $\hat T \equiv T(v=q^\rho)$.
Now at $v=q^\rho$ the free energy \eqref{eq:f_app} is non-differentiable, making $\hat T$
a candidate for $T_c$. Suppose that $T_c=\hat T$.
Then, $\phi(T_c)>  -1/\rho$, i.e.,
the free energy at the critical point is larger than the ground state energy
which if of course impossible.
This means that an entropy driven term generated by a macroscopic number of holes
must be added to \eqref{eq:f_app},
making the assumption that holes are absent false and yields 
\begin{equation}
T_c\geq \hat T\; .
\end{equation}

\section{Eq. (10) is lattice independent}
\label{app3}
We generalize Eq. \eqref{eq:e/k}
\begin{equation}
\label{eq:b/k}
b/k \leq 2\delta+o(1)\; ,
\end{equation} 
obtained for the honeycomb lattice, to the triangular and square lattices.
The notations $k,m$ and $b$ refer to  
the number of faces, sites and boundary sites, respectively, of an arbitrary beast.
The derivations are valid for beasts with no holes.

\subsubsection{Triangular lattice}
\label{sub:triangular}
The total number of sites of any cluster on the triangular lattice can be decomposed 
into (non-zero) numbers $n_1,n_2,n_3,n_4,n_6$ of sites belonging to 
one, two, three, four, six faces, respectively (it is impossible to uniformly colour five faces with a joint vertex, as the 
face-interaction imposes that the sixth face sharing that vertex will have the same colour.
Therefore $n_5=0$).
Similarly to\ (10) in the main text we write
\begin{eqnarray}
\label{eq:n_12346}
n_1 + 2n_2 + 3n_3 + 4n_4 + 6n_6 &=& 3k\; ,\nonumber\\
n_1 + n_2 + n_3 + n_4 +n_6 &=& m\; ,\nonumber\\
n_1 + n_2 + n_3 + n_4 &=& b\; .
\end{eqnarray}
Noticing that the minimal asymptotic number of sites per face on the triangular
lattice is $\rho=1/2$ and expressing $b/k$ and $\delta+o(1)=m/k-1/2$
in terms of $n_1,...,n_6$, we have that \ (\ref{eq:b/k}) immediately follows if and only if
\begin{equation}
\label{eq:constraint_triangular}
2n_1+n_2\geq n_4\; .
\end{equation}
In order to prove\ (\ref{eq:constraint_triangular}) we 
first consider a single animal $a$.
We notice that one gains a total curvature of $2\pi$ when travelling in six different directions along
the animal's perimeter. Non-zero contributions are attributed to crossing $n_{1a},n_{2a}$ and $n_{4a}$
\footnote{The quantities $n_{1a},n_{2a}$ and $n_{4a}$ correspond to the number of sites
of animal $a$
belonging to one, two and four faces, respectively.}
 vertices. A vertex of each type has 
a curvature of $2\pi/3,\pi/3$ and $-\pi/3$, respectively. We thus have   
\begin{equation}
\label{eq:curvature_triangular}
2n_{1a}+n_{2a}-n_{4a}=6\; .
\end{equation}
We next observe that a beast is essentially made of 
$s$ animals interacting via $s-1$ vertices.
Thus, the total number of sites of the beast, belonging to one, two and four faces is
\begin{eqnarray}
\label{eq:n_124_beasts}
n_1&=&\sum_a n_{1a}-2s+2\; ,\nonumber\\ 
n_2&=&\sum_a n_{2a}+s-1\; ,\nonumber\\
n_4&=&\sum_a n_{4a}\; ,
\end{eqnarray}
respectively.
Summing over\ (\ref{eq:curvature_triangular}) and using\ (\ref{eq:n_124_beasts}) lead to
\begin{equation}
2n_1+n_2-n_4=3s+3\; ,
\end{equation}
which completes the proof.

\subsubsection{Square lattice}
Let $n_1,n_2,n_3$ and $n_4$ be the number of sites belonging to 
one, two, three and four faces, respectively, of a given cluster on
the square lattice.
Then 
\begin{eqnarray}
\label{eq:n_1234}
n_1 + 2n_2 + 3n_3 + 4n_4 &=& 4k\; ,\nonumber\\
n_1 + n_2 + n_3 + n_4 &=& m\; ,\nonumber\\
n_1 + n_2 + n_3 &=& b\; .
\end{eqnarray}
On the square lattice there are $\rho=1$ sites per face for simple large clusters.
Writing $\delta+o(1)=m/k-1$ and $b/k$ by means of $n_1,...,n_4$ results in\ (\ref{eq:b/k}), iff \begin{equation}
\label{eq:constraint_square}
n_1\geq n_3
\end{equation}
holds.
To prove\ (\ref{eq:constraint_square}) we, again, first consider a single animal $a$.
We see that any vertex on its perimeter with a non-zero curvature, belongs either to a single face or to three faces. A vertex of each type contributes $\pi/2$ and $-\pi/2$, respectively, to the total curvature of $2\pi$.
Thus, 
\begin{equation}
\label{eq:curvature_square}
n_{1a}-n_{3a}=4\; .
\end{equation}
Next we consider a beast with which can be decomposed
into $s$ animals. 
Employing procedures similar to those described for the triangular lattice,
we wind up with
\begin{equation}
n_1-n_3=2s+2\; ,
\end{equation}
which completes the proof.


\section{Energy-dependent observables and magnetization}
\label{app2}
 
\begin{figure}[htb!!]
\includegraphics[width = \columnwidth]
{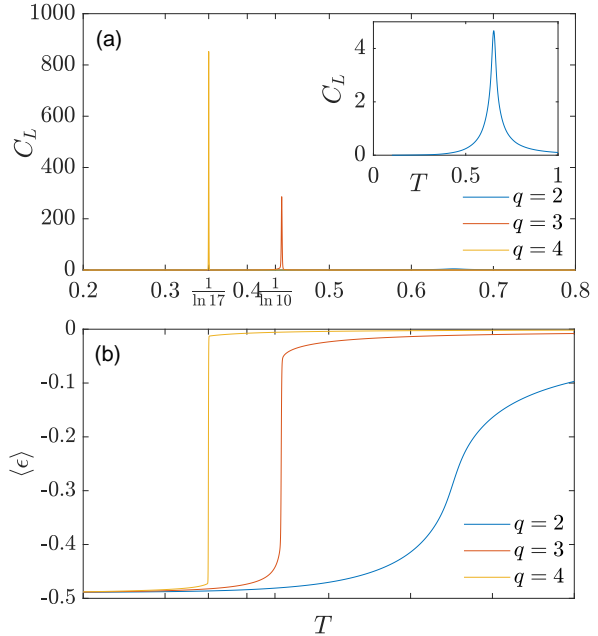}
\caption{\label{fig:A1}
Energy PDF first and second moment-dependent observables as a function of the temperature
for $q=2,3,4$ and $L=45$. The bound $1/\ln (q^2+1)$ on the first order critical temperature is indicated in the cases of $q=3$ and $q=4$.
(a) The specific heat $C_L$. A blow-up in the case of $q=2$ is given in the inset.
(b) The internal energy $\langle\epsilon\rangle$.}
\end{figure}

We discuss additional numerical manifestations of the occurrence and nature
of phase transitions on the honeycomb lattice.
Some of them are remarkably captured by the 
specific heat and internal energy which are closely related to 
the first and second moments, respectively, of the energy PDF.
The specific heat per spin is given by
\begin{equation}
c_L = L^{d}\beta^2(\langle\epsilon^2\rangle-\langle\epsilon\rangle^2)\; ,
\end{equation}
and the internal energy is $\langle\epsilon\rangle$.
The thermal averages $\langle ...\rangle$ are taken with respect 
to the PDF \footnote{In practice we calculate moments of the distribution associated with $\Omega(E)$,
$\frac{\sum_E E^n\Omega(E)e^{-\beta E}}{\sum_E \Omega(E)e^{-\beta E}}$
with $E=L^d\epsilon$, and rescale them properly.}
\begin{equation}
p_L(\epsilon) \propto g_L(\epsilon)e^{-\beta L^d\epsilon}\; ,
\end{equation}
and $g_L(\epsilon)$ is the density of states with energy $\epsilon$.
Plots of the two observables are given in Fig.\ \ref{fig:A1}.
Fig.\ \ref{fig:A1}(a) shows the variation of the specific heat with temperature.
While a clear sharp and narrow peak is observed at $q=3$ and $q=4$ (first order),
a broad, two order of magnitude smaller (essentially hardly noticed on a uniform scale) peak,
is seen when $q=2$. The latter scenario is indeed typical to second order
transitions where large energy fluctuations are present on a relatively
large energy scale. In Fig.\ \ref{fig:A1}(b) we plot the internal energy against the temperature.
A latent heat proportional to the ground state energy (per spin) of one-half,
is evident for $q=3,4$. In the Ising-like case of $q=2$, however, 
the internal energy is a moderate monotonically increasing function of the temperature.
Note also the nice proximity of the positions of both the peak and the jump in energy,
to the approximated critical temperature $T_c\approx 1/(\ln q^\rho+1)$ (with $\rho=2$) for $q=3$
and, in particular, for $q=4$.

Another observable which presents a typical first order behavior 
when computed for the $q=4$ model is 
the magnetization at time (Monte-Carlo sweep) $t$ , given by
\begin{equation}
\label{eq:magnetization}
m(t) = \frac{qx(t)-1}{q-1}\; ,
\end{equation}
where $x(t)$ is the maximal fraction of spins which are simultaneously at the same Potts state.
Indeed, since $q^{-1}\leq x(t)\leq 1$,\ (\ref{eq:magnetization}) implies that $0\leq m(t)\leq 1$. 
We employ the Metropolis  
method to measure the magnetization\ (\ref{eq:magnetization}).
We also simulate the energy density according to\ (1).

\begin{figure}[h]
\begin{center}
\includegraphics[width = \columnwidth,height = 7cm]{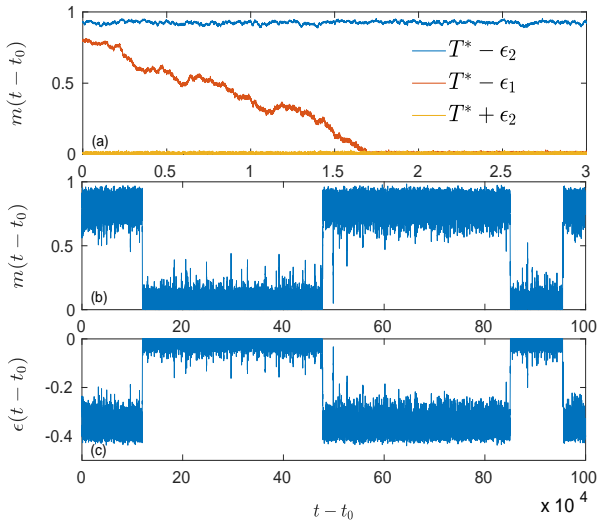}
\end{center}
\caption{\label{fig:A2}
Magnetization and energy density against time for the $q=4$ model on the honeycomb lattice,
simulated at the vicinity of $T^\ast = (2\ln 4)^{-1}$. Two different samples with $L=100$ and $L=20$ are used.
We let the system equilibrate starting from a totally ordered configuration,
and translate the time frame by the equilibration time $t_0$.
(a) $L=100$, $t_0 = 2000$, $\epsilon_1 = 0.001,\ \epsilon_2=0.01$ 
(b) $L=20$, $t_0 = 1000$, $T = T^\ast-0.025$ 
(c) $L=20$, $t_0 = 1000$, $T = T^\ast-0.025$ .}  
\end{figure}

In Fig.\ \ref{fig:A2}(a) we plot the magnetization for a large sample
($L=100$) at the vicinity of $T^\ast$ for $q=4$.
Note that $T^\ast-\hat T=(\ln 16)^{-1}-(\ln 17)^{-1}\approx 0.008$
so it is reasonable that the observed metastability at $T^\ast-\epsilon_1$ 
agrees with the expected metastability at $T_L\approx T^\ast-\epsilon_1\approx T_c$ satisfying
the usual first order relation $|T_c-T_L|\propto L^{-d}$. 
When distorted in $\epsilon_2> \epsilon_1$ below (above) $T^\ast$, or, at temperatures smaller (larger) than $T_c$, the system rapidly relaxes to the ordered (disordered) state, respectively.
Since relaxation times for large samples, when approaching $T_c$, are extremely long,
we additionally simulate a small sample ($L=20$) near $T_c$ and plot the time dependent magnetization and energy density, in Fig.\ \ref{fig:A2}(b) and Fig.\ \ref{fig:A2}(c). As clearly observed in these figures, the large fluctuations enable the system to visit coexisting states in a reasonable time.



%

\end{document}